# A Theoretical Framework for the Most Probable Distribution of Meta-structures in Materials

Wenhao He[1,2], Zhibin Lu[1,2,*]

[1] *State Key Laboratory of Solid Lubrication, Lanzhou Institute of Chemical Physics, Chinese Academy of Sciences, Lanzhou 730000, China*

[2] *Scientific Data Center, Lanzhou Institute of Chemical Physics, Chinese Academy of Sciences, Lanzhou 730000, China*

\* Corresponding author. E-mail: zblu@licp.cas.cn

**Abstract**

Inspired by the principle of equal probability proposed by Boltzmann in the 1870s, we establish a theoretical framework for the most probable distribution of meta-structures in materials. Furthermore, we validate the reliability of this theoretical framework based on statistical results of these meta-structures within randomly generated binary alloys. Finally, combining this theoretical framework, the high-throughput first-principles computations and machine learning, we determine the atomic chemical potentials in the binary FeCr alloy, thereby providing a demonstrative application. This theoretical framework will open a new research area and lay a foundation for the atomic-scale design of targeted properties.

The perspective that microscopic structures determine properties has achieved tremendous success in material design [1,2]. Hence, there has been a persistent curiosity to establish the structure-property relationships in the field of materials science [3-12]. Without a doubt, the quantitative characterization of microstructures is a prerequisite to decipher structure-property relationships. The current characterizations of microstructures mainly focused on phase, component content, degree of disorder [10], chemical short-range order [13] and so on. And, there is a lack of accurately direct methods to quantify microstructure using atomic arrangements. With the rise of doping techniques for modification [14] and high-entropy alloys [15], however, the atom-arrangements urgently need to be explored, as it is crucial for the proactive design of targeted properties at the atomic scale, such as the catalytic efficiency [16], piezoelectricity [17], friction [18], etc. Unfortunately, this issue has not garnered the attention deserved due to the inherent complexity of atomic arrangements and the difficulties associated with experimental detection. Despite the current reports of some methods for directly imaging atomic structures [13,19,20], it is still very difficult to accurately determine atomic types from images of high-resolution transmission electron microscopy (HRTEM) and atomic electron tomography (AET) due to the inability to distinguish atoms with similar atomic numbers.

In the current immature experimental conditions, establishing a theoretical framework for atomic distribution is the first option. The assumption of equal probability principle (EPP) proposed by Boltzmann in the 1870s may be helpful. In analogy to the pivotal roles played by the postulate of the constancy of the light speed in special relativity [21] and the hypothesis of energy quantization in the development of quantum mechanics [22], the EPP laid the foundation for the statistical physics. This principle can be stated that all possible microscopic states of the system are assumed to occur with equal probability in an isolated system at thermodynamic equilibrium. Based on it, the most probable distributions of identical particles, fermions and bosons, given macroscopic quantities such as particle number, volume and energy, have been derived, namely the Boltzmann distribution, Fermi distribution, and Bose distribution. Recently, the EPP has also been introduced into wider fields [23].

In this study, we introduce the EPP into the field of materials science to investigate the distribution of atom-arrangements in a binary infinite solid solution. Firstly, we quantify the random arrangement of atoms through the concept of meta-structures, where the selection of these meta-structures must effectively characterize the fundamental features of the atomic arrangement. Subsequently, establishing the relationship between the number of meta-structure arrangements, which are defined as states, and their distributions, given specific atom numbers, atomic compositions and phases. Following this, inspired by the principle of equal probability, that is each state occurs with equal likelihood, we postulate that meta-structure adheres to the most probable distribution with the maximum number of states. Adhering to this methodology, we construct a theoretical framework for the most probable distribution of meta-structures (MPDM).

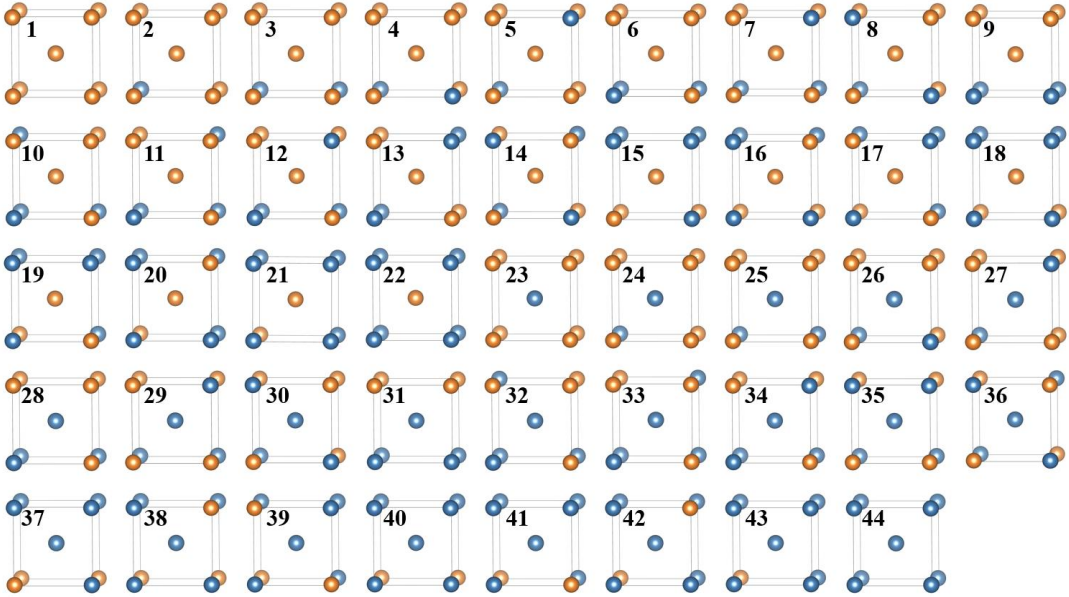

FIG. 1. Illustrations of 44 meta-structures in AB binary alloy with the BCC structure. Brown balls represent A atoms and blue balls represent B atoms.

Taking AB alloy with the body-centered cubic structure (BCC) as an example, there are 44 meta-structures, as shown in Fig. 1. Here, the unit cell is considered as the meta-structure, which can distinguish different atomic arrangements. Table 1 shows the B content $\eta$ in each meta-structure and its degeneracy $\omega$. The degeneracy is determined by symmetry and reflects the number of mirror images of meta-structure.

Assuming the system contains a total of $N$ atoms, where the fractions of A and B atoms are $N_A$ and $N_B$, respectively. The A atoms arrange in the body-centered of 1-22 types of meta-structures, while B atoms arrange in the body-centered of 23-44 types. Assuming that the distribution of meta-structures is $\{a_l\}$. The corresponding number of states $\Omega$ will be defined as

$$\Omega = \frac{N_A!}{\prod_{l=1}^{22} a_l!} * \prod_{l=1}^{22} \omega_l^{a_l} * \frac{N_B!}{\prod_{l=23}^{44} a_l!} * \prod_{l=23}^{44} \omega_l^{a_l} \tag{1}$$

Take the logarithm of Equation (1) and use the $lnN! = N*(lnN-1)$, which holds as $N$ is much greater than 1, to obtain

$$ln\Omega = N_A * lnN_A - \sum_{l=1}^{22} a_l * \ln\left(\frac{a_l}{\omega_l}\right) + N_B * lnN_B - \sum_{l=23}^{44} a_l * \ln\left(\frac{a_l}{\omega_l}\right) \quad (2)$$

In addition, there are three constraints. Firstly, the number of A atoms is equal to the total number of 1-22 meta-structures. Secondly, the number of B atoms is equal to the total number of the 23-44 meta-structures. Thirdly, the number of B atoms is also equal to the total content of B in all meta-structures. The specific expressions are $N_A = \sum_{l=1}^{22} a_l$, $N_B = \sum_{l=23}^{44} a_l$ and $N_B = \sum_{l=1}^{44} \eta_l * a_l$. Solving for the $\{a_l\}$ corresponding to the maximum of $ln\Omega$ is a typical conditional extremum problem, which we tackle by using the Lagrange multiplier method. Let

$$F = ln\Omega - \alpha_1(\sum_{l=1}^{22} a_l - N_A) - \alpha_2(\sum_{l=23}^{44} a_l - N_B) - \beta(\sum_{l=1}^{44} \eta_l * a_l - N_B) \quad (3)$$

where $\alpha_1$, $\alpha_2$ and $\beta$ are the multipliers. Taking the derivative of $F$ with respect to $a_l$ to obtain

$$\frac{\delta F}{\delta a_l} = -\left(\ln\frac{a_l}{\omega_l} + 1\right) - \alpha_1 - \beta * \eta_l \quad (1 \leq l \leq 22) \quad (4)$$

$$\frac{\delta F}{\delta a_l} = -\left(\ln\frac{a_l}{\omega_l} + 1\right) - \alpha_2 - \beta * \eta_l \quad (23 \leq l \leq 44) \quad (5)$$

Setting $\frac{\delta F}{\delta a_l} = 0$ can lead to the distribution of meta-structures $\{a_l\}$ being determined as

$$a_l = \omega_l * exp\left(-1 - \alpha_1 - \beta * \eta_l\right) \quad (1 \leq l \leq 22) \quad (6)$$

$$a_l = \omega_l * exp\left(-1 - \alpha_2 - \beta * \eta_l\right) \quad (23 \leq l \leq 44) \quad (7)$$

By substituting Equations (6, 7) into the three constraint conditions, we can establish a relationship between the B content $c_B$, which defined as $c_B = N_B/N$, and the three multipliers $\alpha_1$, $\alpha_2$ and $\beta$, as shown Equations (9-11). This allows us to solve for the values of the $\alpha_1$, $\alpha_2$ and $\beta$ based on the $c_B$. Further substitute $\alpha_1$, $\alpha_2$ and $\beta$ back into Equations (6, 7) to determine the distribution $\{a_l\}$ of meta-structures for a specific $c_B$. However, an analytical solution for the $\beta$ cannot be derived directly due to the inherent complexity of Equation (9). Therefore, we employ numerical methods for solving Equation (9) to obtain the $\beta$ corresponding to different $c_B$. The detailed algorithms are shown in the Supplementary Materials.

$$c_B = \frac{\frac{1-c_B}{c_B} * \frac{\sum_{l=23}^{44} \omega_l * exp(-\beta * \eta_l)}{\sum_{l=1}^{22} \omega_l * exp(-\beta * \eta_l)} * \sum_{l=1}^{22} \eta_l * \omega_l * exp(\beta * \eta_l) + \sum_{l=23}^{44} \eta_l * \omega_l * exp(-\beta * \eta_l)}{\frac{1-c_B}{c_B} * \frac{\sum_{l=23}^{44} \omega_l * exp(-\beta * \eta_l)}{\sum_{l=1}^{22} \omega_l * exp(-\beta * \eta_l)} * \sum_{l=1}^{22} \omega_l * exp(-\beta * \eta_l) + \sum_{l=23}^{44} \omega_l * exp(-\beta * \eta_l)} \quad (9)$$

$$exp(-1-\alpha_1) = N * (1-c_B)/\sum_{l=1}^{22} \omega_l * exp(-\beta * \eta_l) \quad (10)$$

$$exp(-1-\alpha_2) = N * c_B/\sum_{l=23}^{44} \omega_l * exp(-\beta * \eta_l) \quad (11)$$

Table 1 Content of B atoms $\eta$ and the degeneracy $\omega$ in each meta-structure $l$ of AB alloy with BCC structure.

| $l$ | 1 | 2 | 3 | 4 | 5 | 6 | 7 | 8 | 9 | 10 |
|---|---|---|---|---|---|---|---|---|---|---|
| $\eta$ | 0/16 | 1/16 | 2/16 | 2/16 | 2/16 | 3/16 | 3/16 | 3/16 | 4/16 | 4/16 |
| $\omega$ | 1 | 8 | 12 | 12 | 4 | 24 | 24 | 8 | 6 | 8 |
| $l$ | 11 | 12 | 13 | 14 | 15 | 16 | 17 | 18 | 19 | 20 |
| $\eta$ | 4/16 | 4/16 | 4/16 | 4/16 | 5/16 | 5/16 | 5/16 | 6/16 | 6/16 | 6/16 |
| $\omega$ | 24 | 24 | 6 | 2 | 24 | 24 | 8 | 12 | 12 | 4 |
| $l$ | 21 | 22 | 23 | 24 | 25 | 26 | 27 | 28 | 29 | 30 |
| $\eta$ | 7/16 | 8/16 | 8/16 | 9/16 | 10/16 | 10/16 | 10/16 | 11/16 | 11/16 | 11/16 |
| $\omega$ | 8 | 1 | 1 | 8 | 12 | 12 | 4 | 24 | 24 | 8 |
| $l$ | 31 | 32 | 33 | 34 | 35 | 36 | 37 | 38 | 39 | 40 |
| $\eta$ | 12/16 | 12/16 | 12/16 | 12/16 | 12/16 | 12/16 | 13/16 | 13/16 | 13/16 | 14/16 |
| $\omega$ | 6 | 8 | 24 | 24 | 6 | 2 | 24 | 24 | 8 | 12 |
| $l$ | 41 | 42 | 43 | 44 | | | | | | |
| $\eta$ | 14/16 | 14/16 | 15/16 | 16/16 | | | | | | |
| $\omega$ | 12 | 4 | 8 | 1 | | | | | | |

We further determine whether this distribution $a_l$ corresponds to the maximum number of states. This requires calculating $ln\Omega$ second-order variation

$$\frac{\delta^2 ln\Omega}{\delta a_l^2} = -\frac{1}{a_l} \quad (12)$$

A negative second-order derivative indicates that the distribution is one that maximizes $ln\Omega$.

As mentioned above, the distribution corresponding to Equations (6, 7) is derived based on the maximum number of states, so we refer to this distribution as the most probable distribution. Whether the system is necessarily in the most probable distribution, we also need to know the dependence of the number of states on the distribution. If the state number has an extremely steep peak at the most probable distribution, then the system follows the most probable distribution. To prove this point, we will carry out the second-order expansion of the state number $ln(\Omega + \Delta\Omega)$ corresponding to the deviation $\Delta a_l$ from the most probable distribution.

$$ln(\Omega + \Delta\Omega) \approx ln\Omega + \delta ln\Omega + \frac{1}{2}\delta^2 ln\Omega \tag{13}$$

$$ln\left(\frac{\Omega + \Delta\Omega}{\Omega}\right) \approx -\frac{1}{2}\left(\frac{\Delta a_l}{a_l}\right)^2 a_l \tag{14}$$

Assuming a deviation of $\frac{\Delta a_l}{a_l} = 10^{-5}$ and considering a macroscopic system where $a_l \approx 10^{23}$, we find that $\frac{\Omega + \Delta\Omega}{\Omega} \approx exp(-5 * 10^{12})$. It indicates that even a minute deviation in the distribution leads to a ratio of the number of states corresponding to this deviation and that at the most probable distribution is virtually zero. Therefore, we argue that the distribution of meta-structures follows the most probable distribution.

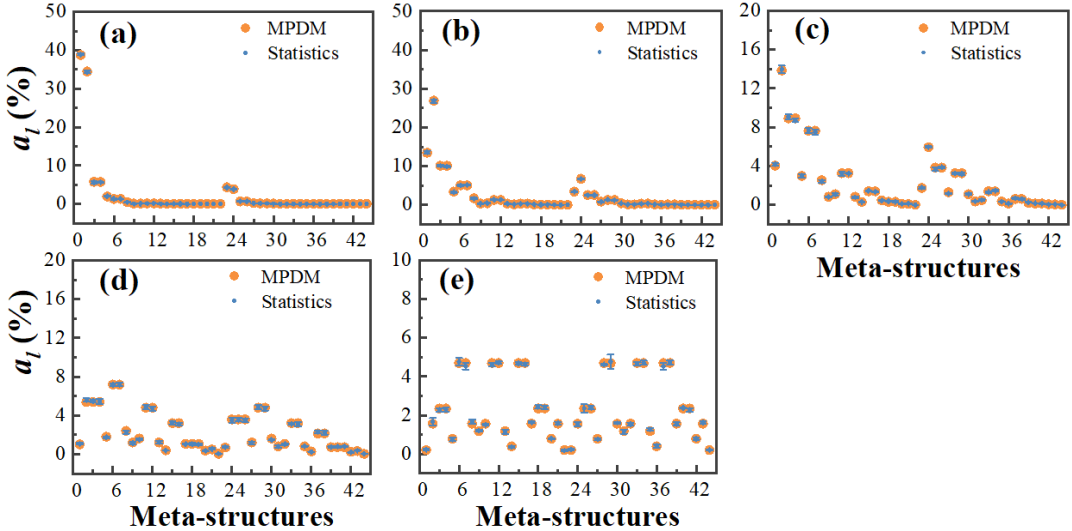

FIG. 2. Distributions of 44 types of meta-structures in the infinite solid solution binary AB alloy with the BCC structure from the theoretical framework for MPDM and statistical results from randomly generated systems. The corresponding B content are 10% (a), 20% (b), 30% (c), 40% (d), and 50% (e), respectively. The statistical results for each B content are derived from five randomly generated AB systems, each containing 16,000 atoms.

We firstly substantiated the validity of the theoretical framework for MPDM by taking an example of the infinite solid solution binary AB alloy with the BCC structure and conducting a rigorous analysis. We considered 5 AB systems for each B content. B contents of 10%, 20%, 30%, 40% and 50% were taken into account. Each AB system contains 16000 atoms, which was randomly constructed. Then counting the meta-structures in these systems and compared them with the most probable distribution from the theoretical framework for MPDM. Random structure generation and statistical methods were provided in detail in the Supplementary Materials. As shown in Fig. 2, the consistency between them indicates the rationality of the theoretical framework for MPDM.

Further, we take the chemical potential of atoms in FeCr alloys with the BCC structure as a brief application of the theoretical framework for MPDM. The atomic chemical potential is intimately related to properties, such as the corrosion performance and tribo-chemical reactions at interfaces. Currently, the classical Widom-type substitution technique [24] and the first-principles were usually combined to calculate the atomic chemical potential in alloys [25-27]. However, this approach is limited to calculating the atomic chemical potential within specific configurations, making it difficult to obtain the statistical average. Furthermore, deriving the variation of atomic chemical potential with atomic concentration is virtually unfeasible using this strategy. In contrast, the theoretical framework for MPDM proposed in this study enables the calculation of the average chemical potential as a function of atomic concentration.

If introducing a Fe atom into the FeCr alloys, it can occupy any one of the body-center of 1-22 types of meta-structures, and the probability of its occupation equal to the distribution concentration of meta-structures. Correspondingly, the introducing Cr atom will occupy any one of body-center of 23-44 types of meta-structure. Therefore, the average chemical potential of Fe atom $\mu_{Fe}$ and Cr atom $\mu_{Cr}$ can be defined as

$$\mu_{Fe} = \sum_{l=1}^{22} a_l * e_l / (1 - c_{Cr}) \tag{12}$$

$$\mu_{Cr} = \sum_{l=23}^{44} a_l * e_l / c_{Cr} \tag{13}$$

where $e_l$ is the energy of the meta-structure.

In order to obtain the energy $e_l$ of each meta-structure, it is critical to obtain the total energy of abundant FeCr systems with different $c_{Cr}$. Here, total 239 FeCr systems were considerd. To save operational time, we have developed a High-throughput First-Principles Calculation and Data Collection Software specifically tailored for computing total energies of FeCr systems and collecting data, which were provided in detail in the Supplementary Materials. The first-principles calculations were carried out with the Vienna Ab initio Simulation Package (VASP) [28,29] based on the density functional theory (DFT) [30,31]. The projected augmented wave (PAW) [32] pseudopotentials were employed in the calculations within the generalized gradient approximation (GGA) with Perdew-Burke-Ernzerhof (PBE) [33] functional for the exchange and correlation energies. A cutoff energy of 300 eV was used for the plane-wave expansion. The internal structure relaxations stopped when the residual force on each atom was less than 0.01 eV/Å. The Monknorst-Pack k-meshes of 4×4×4 was used. Spin polarization was been taken into account. During calculations, both the positions of atoms and volumes of systems were allowed to relax.

After completing the first-principles calculations, further data collection was performed to gather the total energies and the distributions of meta-structures within each FeCr system, as summarized in Table S1. Subsequently, based on these data, we employed a neural network approach to learn the energy of each meta-structure. The detailed machine learning algorithms were shown in the Supplementary Materials. Fig. 3 (a) shows the energies $e_l$ of the meta-structures. Obviously, it decreases with its orders. That is, the meta-structures with high Cr contents have smaller energy. To verify the accuracy of the fitted $e_l$, Fig. 3(c) shows the total energies $E_{Fitting}$ calculated from the fitted $e_l$ and the $E_{DFT}$ from DFT for 239 FeCr systems. All of differences $E_{Error}$ between $E_{Fitting}$ and $E_{DFT}$ are less than 10meV/atom, and most of them are less than 5meV/atom. It indicates that the fitted $e_l$ have high accuracy. Fig. 3(b) shows the chemical potential of Fe atom $\mu_{Fe}$ and Cr atom $\mu_{Cr}$ in FeCr alloys. Both $\mu_{Fe}$ and $\mu_{Cr}$ decrease with the increasing $c_{Cr}$.

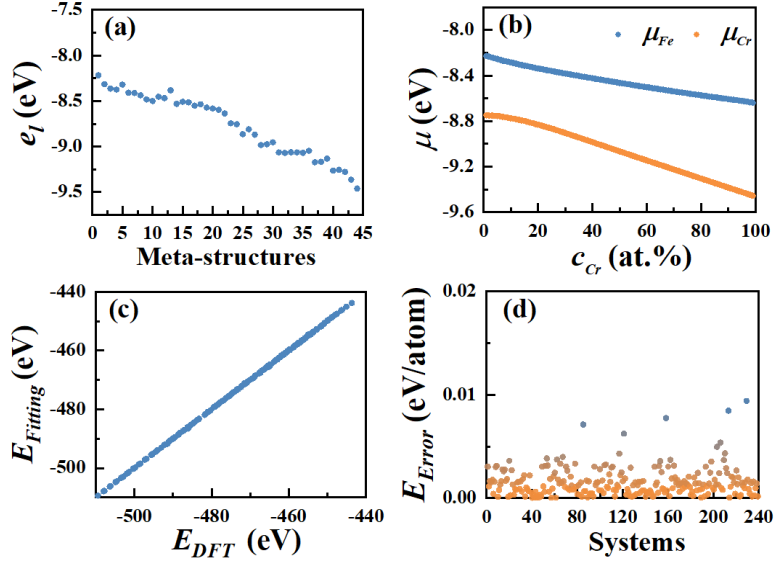

FIG. 3 The fitting energies of the meta-structures by using the machine learning method and the total energies of 239 FeCr systems from DFT calculations (a). The average chemical potential of Fe atom $\mu_{Fe}$ and Cr atom $\mu_{Cr}$ in FeCr alloys vs. Cr contents $c_{Cr}$ (b). The fitting total energies $E_{Fitting}$ vs. the $E_{DFT}$ from DFT calculations for 239 FeCr systems (c). The differences $E_{Error}$ between $E_{Fitting}$ and $E_{DFT}$ (d). All FeCr systems, consisting of 54 atoms each, are randomly constructed, with the Cr content uniformly distributed across a range spanning from 0 to 100%.

Although the theoretical framework for MPDM was originally derived for binary alloys, with a reasonable definition of meta-structure, it may be applicable to multi-component alloys and various compound systems as well. Notably, prior to its derivation, this theoretical framework relies on an important assumption, namely the equiprobability principle, which posits that all possible arrangements, that is states, of the meta-structures have equal probabilities. This assumption may not always hold, as it solely embodies the principle that microscopic structures tend towards greater disorder. However, the evolution of microstructures is governed not only by disorder but also by the principle of energy minimization. In other words, meta-structures with lower energies may be more abundant than predicted by this framework, which will resulted in the common phenomenon of segregation in alloys [13,34,35]. Nonetheless, this theoretical framework retains significant importance. We will discuss its significance from three perspectives. Firstly, at high temperatures, entropy dominates the energy to the system, rendering the system in a state of maximum disorder [36]. If a material is prepared through rapid cooling, the high-temperature microstructure is preserved, rendering the theoretical framework directly applicable. Secondly, for some alloys, segregation occurs only when the content of a certain element reaches a critical level. For such materials, this theoretical framework is applicable within the range of element contents that correspond to a disordered state. Lastly, when energy minimization becomes a crucial factor, as in the case of segregation phenomena, we should introduce this constraint into the framework and develop a theory of most probable distribution that incorporates energy. This may be an important direction for future development of this theoretical framework.

In conclusion, we establish a theoretical framework for the most probable distribution of meta-structures in the infinite solid solution binary alloy, which is inspired by the principle of equal probability proposed by Boltzmann in the 1870s. Furthermore, we validate the reliability of this theoretical framework based on statistical results of these meta-structures within randomly generated binary alloys. Finally, combining the high-throughput first-principles computations and machine learning, we successfully determine energies of all meta-structures within FeCr alloy. And obtaining the average chemical potentials of Fe and Cr in FeCr alloy with any Cr concentrations based on the theoretical framework, thereby providing a demonstrative application.

## Acknowledgements

This work is supported by the Strategic Priority Research Program of the Chinese Academy of Sciences (Grant No. XDB 0470203), National Natural Science Foundation of China (No. 12302128), West Light Foundation of The Chinese Academy of Sciences (xbzg-zdsys-202305), the CAS Project for Young Scientists in Basic Research (YSBR-023) and LICP Cooperation Foundation for Young Scholars (HZJJ23-4).